\documentclass[lettersize,journal]{IEEEtran}
\usepackage{amsmath,amsfonts}
\usepackage{algorithmic}
\usepackage{algorithm}
\usepackage{array}
\usepackage[caption=false,font=normalsize,labelfont=sf,textfont=sf]{subfig}
\usepackage{textcomp}
\usepackage{stfloats}
\usepackage{url}
\usepackage[bookmarks=false]{hyperref}
\usepackage{verbatim}
\usepackage{graphicx}
\usepackage{multirow}
\usepackage{booktabs}
\usepackage{cite}
\begin{document}

\title{ Spatial Graph Convolutional Neural Network via Structured Subdomain  Adaptation and Domain Adversarial Learning for Bearing Fault Diagnosis}

\author{Mohammadreza Ghorvei, Mohammadreza Kavianpour, Mohammad TH Beheshti*, Amin Ramezani \\
Department of Electrical and Computer Engineering\\Tarbiat Modares University\\ Tehran, Iran\\
E-mail:m.rezaghorvei@ieee.org
\thanks{\textbf{This work has been submitted to the IEEE Transactions on Instrumentation and Measurement for possible publication.\\ Copyright may be transferred without notice, after which this version may no longer be accessible.}}
}

\maketitle

\begin{abstract}
Unsupervised domain adaptation (UDA) has shown remarkable results in bearing fault diagnosis under changing working conditions in recent years. However, most UDA methods do not consider the geometric structure of the data. Furthermore, the global domain adaptation technique is commonly applied, which ignores the relation between subdomains. This paper addresses mentioned challenges by presenting the novel deep subdomain adaptation graph convolution neural network (DSAGCN), which has two key characteristics: First, graph convolution neural network (GCNN) is employed to model the structure of data. Second, adversarial domain adaptation and local maximum mean discrepancy (LMMD) methods are applied concurrently to align the subdomain's distribution and reduce structure discrepancy between relevant subdomains and global domains. CWRU and Paderborn bearing datasets are used to validate the DSAGCN method's efficiency and superiority between comparison models. The experimental results demonstrate the significance of aligning structured subdomains along with domain adaptation methods to obtain an accurate data-driven model in unsupervised fault diagnosis.
\end{abstract}

\begin{IEEEkeywords}
Unsupervised fault diagnosis, graph convolution neural network, subdomain adaptation, adversarial domain adaptation
\end{IEEEkeywords}

\section{Introduction}
\IEEEPARstart{R}{olling} bearings, the most important mechanical components of rotary machines, are prone to fault for various reasons, like a harsh working environment and a long working period. The faulty bearing may harm mechanical equipment, which leads to catastrophic accidents. Accordingly, an accurate and effective diagnosis of bearing faults is essential for equipment’s reliable operation\cite{wang2019new}.\\ 
With the advancement of technology, many approaches for intelligent bearing fault diagnosis have been developed. Machine learning is one of them, in which extracted features from pre-processed data are fed into classifiers such as Support Vector Machine (SVM)\cite{wang2021modified}, k-Nearest Neighbors (KNN)\cite{lu2021enhanced}, and Random Forest (RF)\cite{wei2021intelligent} to classify fault types. The prerequisite of accurate predicting is when extracted features are considerably knowledgeable\cite{r1,r2}. There are still drawbacks when there is a particular need for technical expertise in graph feature extraction and feature selection and a low capacity to learn non-linearity and complexity in the data patterns.\cite{lei2020applications}.\\
Deep learning (DL) techniques have been employed in recent years to address the shortcomings of traditional machine learning methods. This technique has more dependable performance in feature learning and extracting more abstract features. Furthermore, no prior knowledge or experience is required for feature engineering by using end-to-end learning algorithms\cite{r3,r4}. Convolution neural network (CNN) \cite{liu2020multitask}, deep auto-encoder (DAE) \cite{liu2020stacked}, deep belief network (DBN) \cite{xing2020distribution}, and recurrent neural network (RNN) \cite{ravikumar2021gearbox} are the most commonly utilized DL techniques in intelligent fault diagnosis. A novel discriminant regularizer in DAE has been proposed in \cite{mao2021new}  to diagnose bearing faults. Deep residual CNN is used in \cite{yang2020fault} ,in which the noise impact is decreased by employing a wide kernel in the first layer of convolution. Also, an attempt is made to reduce the gap produced by the distribution discrepancy between source and target data by utilizing the adaptive batch normalization method. Peng et al. \cite{peng2020multibranch} proposed a multiscale CNN to extract short-time and long-time features and feature fusion to enhance model performance for fault diagnosis. Wang et al. \cite{wang2021intelligent} introduced a normalized CNN to identify the fault, and the suggested model's hyperparameters are optimized using particle swarm optimization.\\
If a substantial number of labeled data is available for model training, and the distribution of training and test data is the same, DL models can perform accurately.  In contrast, the collection of labeled data, particularly faulty data, is not practicable in many real-world applications due to time and cost constraints. Consequently, the train and test data may have different probability distributions due to continuously changing operating conditions of rotary machines, resulting in poor performance and limited generalization of the DL models\cite{me}. Hence, it is essential to develop a strategy that alleviates the gap between training and test data distribution and enhances performance without creating a novel model for new unlabeled data. Over the past few years, the UDA technique has been utilized as a distinctive form of transfer learning in intelligent fault diagnosis, in which learned knowledge from labeled data in the source domain is transferred to the unlabeled data of the target domain through discovering domain-invariant and discriminative characteristics\cite{xu2021ifds}.
Source and target domain are utilized in UDA approaches to training a model with shared weights, which can benefit from both. Moreover, it aims to learn the invariant characteristics across two domains. Coral\cite{c}, is one of the most effective UDA methods that has been widely utilized in unsupervised bearing fault diagnosis recently. This regularization method aims to align sources and targets’ second-order statistics (batch covariances) with a linear transformation. An extended version of coral named deep coral is intergraded with a deep neural network that aims to learn a nonlinear transformation to align layer activations’ correlations\cite{dc}. RMCA-1DCNN model based on Riemann metric correlation alignment loss is proposed in  \cite{cia} to obtain unsupervised fault diagnosis  with domain-invariant and fault-discriminative capacity. Furthermore, UDA follows to reduce the difference distance between distributions in latent space resulting from criteria such as the maximum mean discrepancy (MMD)  and Multi kernel maximum mean discrepancy (MK-MMD) \cite{9146579
,zhang2021joint}. For instance, the MK-MMD is employed in \cite{an2020deep} to decrease the distribution disparity caused by changing operating conditions. Lu et al. \cite{lu2021new} introduced multi-layer MMD to match the distribution between source and target. The objective of these approaches is to decrease the difference between the mean values of the distribution in the two domains, but data characteristics like median and the standard deviation in domains may not be the same after using MMD-based approach \cite{mao2020new}. This reduces the accuracy of classification in the target domain. It should be noticed that if the distribution discrepancy between the data is substantial, some critical data characteristics may be lost while projecting data from two domains to the same feature space. In a nutshell, applying the MMD-based approach is not sufficient individually to learn invariant features in two domains.\\
The discriminative adversarial network for domain adaptation has been introduced with promising outcomes\cite{zhang2021conditional}. The source and target domain distributions are aligned in this technique until the discriminator can determine whether the input data belongs to the source or target domain. Mao et al. \cite{mao2020new} proposed adversarial domain training to identify faults under various working conditions. Xu et al. \cite{xu2021intelligent} employed multi-layer adversarial learning to adjust the domain and boost the model's generalizability. Li et al. \cite{li2020intelligent} proposed two feature extractors, one of which uses MMD techniques and the other adversarial domain training to extract invariant features, ensemble learning is used to improve the accuracy of fault diagnosis. Zhang et al. \cite{zhang2019deep} introduced the multi-adversarial domain adaption approach with the Wasserstein distance criteria to aid in diagnosing bearing faults under various operating conditions. 
The distribution discrepancy between the two domains in bearing fault diagnosis is considerable due to multiple factors such as changing operating conditions and environmental noise. Appropriately, if the distribution difference between the two domains increases, the convergence of adversarial techniques to a stable position will confront difficulties. Hence, providing an effective solution is critical to increasing the stability of adversarial techniques in bearing diagnosis. The domain label, the label for each category, and the data structure as three critical kinds of information represent an essential role in UDA techniques for bridging knowledge transfer from the source to the target domain\cite{ma2019gcan}. The domain label is used to train a domain classifier to represent the global distribution of both domains in adversarial UDA. If the label of each category for the source and target domain data samples is the same, it is expected to be mapped to the same feature space. The data structure comprises the data's intrinsic characteristics, such as probability distribution and geometric structure of data. All three types of information can assist in alleviating the gap between the two domains under different operation conditions and improve the performance of the fault diagnosis model\cite{li2021domain}. As we know, most studies cover just one or two of these types of information and pays inconsiderable regard to the geometric structure of the source and target data. The data structure is  considered as a grid in the mentioned models, which may be a limiting factor for the model's generalization. On the  other side, If we offer data with a graph-based structure as an input to a network like CNN, the desired results are not reached because these networks are built on grid-structured data. Furthermore, the domain adaptation method's objective is to match the distribution of source and target domains. While each domain has multiple different subdomains, the UDA technique does not follow each subdomain's adjusting distribution with the corresponding class. After adjusting the global distribution of two domains in the UDA, some unrelated data from different classes may be nearby to the data of a specific class in latent space, resulting in a decrease in domain adaptability and diagnostic accuracy. One solution to this problem is the LMMD technique \cite{zhu2020deep}, which is an extension of the MMD method. In addition to adjusting the global distribution between domains, the LMMD technique brings the distributions of each identical class as a subdomain in two separate domains closer together in latent space and adapts them.
\\
This paper proposes a unique DSAGCN as a graph convolution neural network (GCNN)-based \cite{gcnn} solution and distribution discrepancy reduction employing LMMD and adversarial loss function to satisfy all of the limitations mentioned earlier. A CNN is used in DSAGCN to extract features from the vibration signal. The extracted features are input into several topology adaptive graph convolutional network (TAGCN) \cite{du2017topology} blocks of learning the geometric structure of the data, which are produced by investigating the relationship between the structural characteristics of the data and propagating structural information in the graph network's parameters. Then, the LMMD loss function and the adversarial loss function jointly alleviate the structure discrepancy in domains distribution. Both of these criteria have advantages that have been described earlier. As a result, the structural information of the data is examined by employing feature modeling in the form of graphs. The classifier also models class labels, and the adversarial domain discriminator distinguishes the domain label for each data. Therefore, the suggested DSAGCN method incorporates all three elements that significantly increase UDA performance.\\

The following are the major contributions of this study:
\begin{enumerate}
\item This paper offers a practical and comprehensive end-to-end DSAGCN method to diagnose cross-domain bearing faults based on graph and domain adversarial discriminator and structured subdomain adaptation. In this proposed method, all information, including data structure, domain label, and labels of each class, are used in an integrated manner to minimize the distribution difference across domains and the distribution difference between relevant subdomains in latent space.
\item The GCN model proposed for this study is the TAGCN method. The simulation results indicate that this model can provide acceptable results by employing graphs with second-order polynomials. Combining with CNN increases the model's effectiveness in identifying bearing faults under different operation conditions.
\item Due to the substantial distribution difference between the source and target domain through changing load while collecting the vibration data, LMMD and an adversarial domain discriminator were employed concurrently to find domain-invariant and discriminative features that would reduce the distribution difference between the two domains and align them. The ablation study demonstrates that the LMMD applied in DSAGCN outperforms coral and other MMD-based loss functions in average accuracy and convergence speed.
\end{enumerate}
The rest of this paper is structured as follows:
The fundamental theory of the approaches utilized in DSAGCN is explained in \hyperref[sec2]{Section \ref{sec2}}. \hyperref[sec3]{Section \ref{sec3}} describes the structure and characteristics of the proposed DSAGCN method. In \hyperref[sec4]{Section \ref{sec4}}, after introducing the dataset used to evaluate the model, the experimental results of the proposed technique are compared with other comparative methods to assess the benefits of the DSAGCN method. Finally, \hyperref[sec6]{Section \ref{sec6}} provides the conclusion.
\section{PRELIMINARIES}  \label{sec2}
\subsection{Graph Convolutional Neural network}
\begin{figure}[t]
    \centering
    \includegraphics[width=\columnwidth]{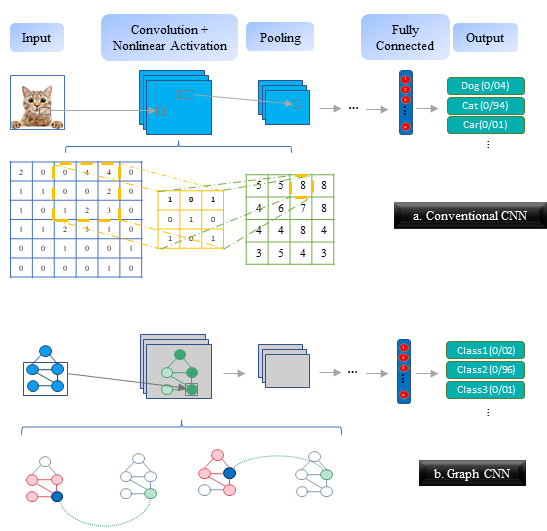}
    \caption{Comparison of conventional CNN and GCNN architecture}
    \label{gcnn}
    \end{figure}

CCNs have essential properties such as local connection, weight sharing, and invariance with shifts, which allow them to be used in various disciplines such as image processing, fault diagnosis, and data with uniform and grid-based structures \cite{chen2021multiscale}.
However, Conventional CNN methods cannot achieve an accurate result in many cases, such as biological systems, social networking applications, and fields where the data has a non-Euclidean and irregular structure\cite{cheung2020graph}. Graph structure outperforms conventional CNNs in these applications. Furthermore, knowing the geometric structure of the data improves model learning and lowers the distribution gap across domains in transfer learning fields. Graph signal processing (GSP) \cite{ortega2018graph} is used to alter conventional CNNs by leveraging graph theory, which provides both a general framework and a rigorous view for GCNN. A conventional CNN and a GCNN are compared in Fig. \ref{gcnn}. Conspicuously, The convolution operator computes the sum of pointwise multiplications of a subset of input by the kernel of that layer. This process is repeated until the kernel has moved through the input. Fig. \ref{gcnn} demonstrates the GCNN architecture and how it differs from a conventional CNN for inputs with irregular structures. A 1-degree filter is shown, with the blue vertex collecting information about the red adjacent vertices A 1-degree filter is shown, with the blue vertex collecting information about the red adjacent vertices.
A graph is commonly expressed as $ G = \left( {\nu ,\varepsilon ,A} \right) $, where $\nu$ is the set of N vertices, $\varepsilon$ denotes the set of edges connecting the two corresponding vertices, and $A$ expresses the adjacency matrix that shows how each vertex is related to the other vertices. Each ${{A_{i,j}}}$ term represents the weight of the connected edge between i and j. The graph Laplacian is defined as $L=D-A$, where L is the graph Laplacian and $D=diag[\sum\nolimits_j {{A_{i,j}}} ]$ is the degree matrix of $A$ \cite{yu2021fault}. Every graph has the option of being directed or undirected. The adjacency matrix is symmetric in an undirected graph since the path between the two vertices is common, but it can be asymmetric in a directed graph due to the path's direction. Any graph can either be weighted or unweighted. GCNN techniques are broadly classified into two types: spectral domain and spatial domain \cite{li2021fault}. spectral domain technique utilizes the Fourier transform of the graph and the generalized Laplacian operator in the spectral approach. The spatial domain approach does not employ the Fourier transform, instead of relying on GSP's definition of graph convolution and the concept of graph shift \cite{sandryhaila2013discrete}. This concept describes how to propagate information from one node to neighboring nodes and replace each node's signal value with a linear combination of signal values in that node's neighborhood. Shift operator is a vital component in GSP. For example, the adjacency matrix $A$ as a shift operator and $x$ as a graph signal, the one-stage propagate is the new graph signal $Ax$, and the n-stage propagate is the signal ${A^n}x$.\\
TAGCN technique \cite{du2017topology} is one of the most acceptable spatial-based methods in GCNNs. A graph convolutional layer is generated in this technique by integrating the concepts of graph convolution and graph shift. The operation of graph convolution on the L-layer convolutional is described without loss of generality. Assume that each vertex of the graph has the ${P_L}$ feature input in this layer, and the vector $x_p^{(L)} \in {\mathbb{R}^{{N_L}}}$ contains the $L$-th layer's input data for the $p$-th feature of all vertices. The graph $G$ indexes the elements  $x_p^{(L)} \in {\mathbb{R}^{{N_L}}}$. Assume that $G_{p,f}^{(L)} \in {\mathbb{R}^{{N_L} \times {N_L}}}$ represents the GCNN's $f$-th filter in the $L$-th layer. Multiplying the $G_{p,f}^{(L)}x_p^{(L)}$ matrix produces the graph convolution. The $f$-th output of the convolution is calculated using the following equation:
\begin{figure}[t]

    \centering
     \includegraphics[width=\columnwidth]{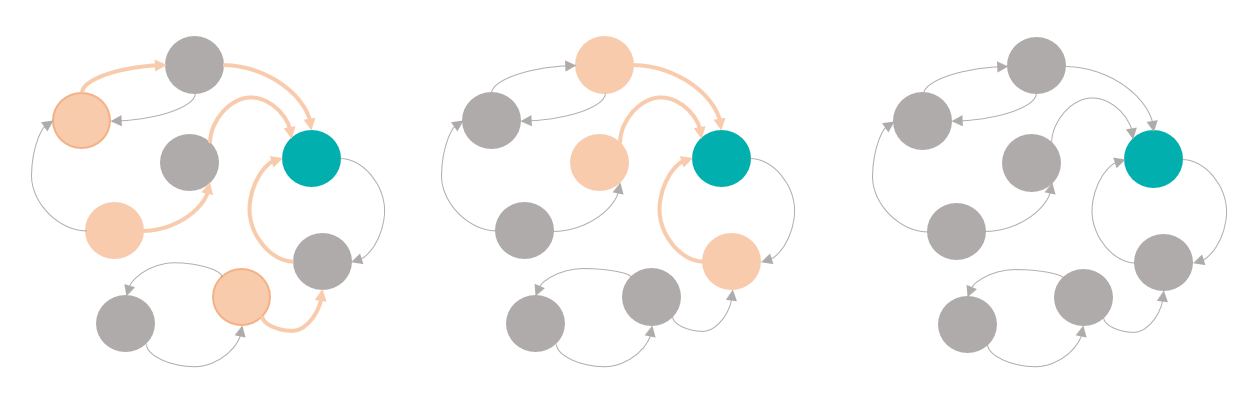}
    \caption{TAGCN polynomial filter for jade color node for $k = 2$. The signal from the orange nodes is propagated to the jade node and aggregated, i.e., ${\alpha _2}{{\bar A}^2}{x^{\left( L \right)}} + {\alpha _1}\bar A{x^{\left( L \right)}} + {\alpha _1}{I_N}{x^{\left( L \right)}}$ }
    \label{TAGCN}
    \end{figure}
\begin{equation}
    \begin{aligned}
y_f^{(L)} = \sum\limits_{p = 1}^{{P_L}} {G_{p,f}^{(L)}x_p^{(L)}}  + b_f^{(L)}{1_{{N_L}}}
\end{aligned}
\end{equation}
where $b_f^{(L)}$ is the bias and ${1_{{N_L}}}$ is a square matrix with one element with dimensions equal to the number of graph vertices. $ G_{p,f}^{(L)}$ can be obtained as a polynomial in $A$:\\
\begin{equation}
    \begin{aligned}
   G_{p,f}^{(L)} = \sum\limits_{k = 0}^K {{\alpha _k}{{\bar A}^k}}   
    \end{aligned}
\end{equation}
In this equation, ${{\alpha _k}}$ and ${\bar A}={D^{ - \frac{1}{2}}} A{D^{\frac{1}{2}}}$ are the polynomial coefficients of the graph filter and the normalized adjacency matrix, respectively. Normalizing the adjacency matrix ensures that all eigenvalues are localized inside a unit circle, resulting in computational stability $G$.\\ The TAGCN method as a spatial-based method can be compared with spectrum methods in different aspects. Basically, The TAGCN technique uses a series of learnable node filters with fixed sizes ranging from size 1 to size $K$ to perform graph convolution. Furthermore, it has O(K) learning complexity, similar to conventional CNN algorithms. TAGCN has a reduced computing burden than spectrum-based techniques, which spectrum-based techniques have a more significant computational burden owing to Fourier transform, inverse Fourier transform, and Eigen decomposition. Aside from the reduced computing burden, the TAGCN technique also has lower computational complexity than spectrum methods. Second-order polynomials can produce good results in this approach, whereas 25-order polynomials from the graph adjacency matrix are required in \cite{defferrard2016convolutional}. Fig. \ref{TAGCN} depicts the TAGCN polynomial filter for $k = 2$. The selected $k$ and graph direction indicate the connections for information propagation and collection from the orange nodes to the jade node in Fig. \ref{TAGCN}.
This graph demonstrates the TAGCN approach's decreased computational complexity compared to other methods, particularly the method provided in \cite{defferrard2016convolutional}.  In addition, unlike the study in \cite{kipf2016semi}, no linear estimation is used in the TAGCN, leading to data loss and lower classification accuracy. TAGCN uses GSP theory to construct filters ranging from 1 to $K$ to avoid this problem. The spectral technique is unusable for directed graphs because the spectral technique utilizes shift graph operators such as graph Laplacian, which is only relevant to undirected graphs, and since the graph Laplacian must be positive semidefinite in order to apply this approach, which is only possible with the symmetry of matrix A.
Unlike spectrum-based strategies, each convolutional layer in the TAGCN technique is unique due to variable-size graph convolutional layers. The Fourier transform will not be unique in spectrum methods if the eigenvalue obtained from the Laplacian is repeated. 
\begin{figure}[t]
    \centering
    \includegraphics[width=\columnwidth]{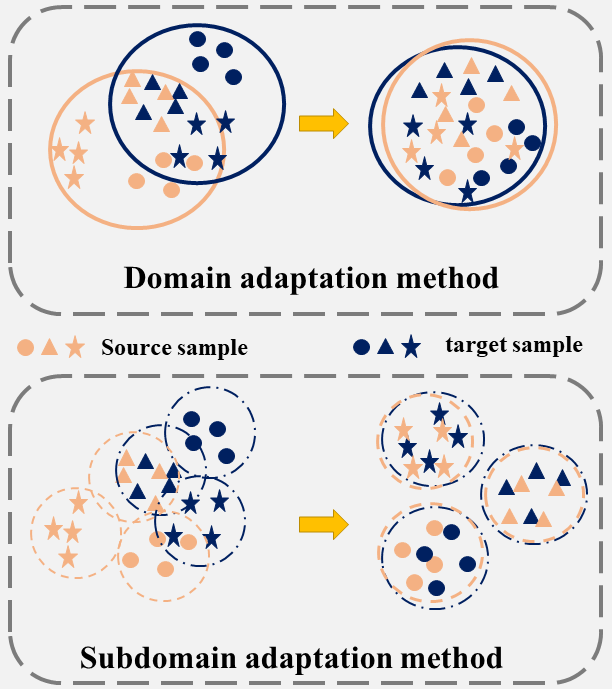}
    \caption{Top: the global DA approach merely reduces the distribution distance across domains. Bottom: The deep subdomain adaptation approach, which takes fine-grained information into account and brings the distribution of subdomains of the same class in two domains closer together.}
\label{subdomain}
    \end{figure}
\subsection{deep structured subdomain adaptation}
DL methods mostly assume that the distribution of test and training data is homogeneous, and if this null hypothesis is false, DL method performance will be  significantly reduced. While the distribution difference between test and training data is seen in many applications. The domain adaptation technique is presented as a solution for this category of issues, in which learning the invariant feature space in the source and target domains leads to a reduction in the difference in domain distribution and an increase in model's performance.
A domain is defined in domain adaptation as $D = \left\{ {\chi ,P\left( X \right)} \right\}$, where $\chi $ and ${P\left( X \right)}$ are the data feature space and their marginal probability distribution, respectively, and $X = \left\{ {{x_i}} \right\}_{i = 1}^n \in \chi $ are instance samples. ${D_S}$ and ${D_T}$ denote the source and target domains, respectively. An unsupervised domain adaptation problem is described as having Labeled source domain  ${D_S} = \left\{ {\left( {x_i^s,y_i^s} \right)} \right\}_{i = 1}^{{n_S}}$ and unlabeled target domain data ${D_T} = \left\{ {x_j^t} \right\}_{j = 1}^{{n_T}}$. The labeled space is denoted by $y$. In this scenario, it is assumed that the feature space, category space, and probabilistic conditional distribution of these two domains are the same, but their probabilistic marginal distribution is different owing to the domain shift. Domain adaptation aims to predict the target domain data labels using transferred information from the source domain \cite{zheng2019cross}.
The global domain adaptation technique's core idea is to roughly adjust the global probability distribution of the source and target domains. Despite the benefits of this technique, some irrelevant data from one class may be close to another one in feature space after aligning distribution, and DA performance suffers due to a lack of relationships between subdomains with the same class in different domains \cite{zhu2020deep}.The subdomain adaptation technique is an efficient solution to avoid this problem. The probability distribution of the same classes in different domains and the discrepancy between them are considered in this method. In addition to aligning the distribution of two domains, this strategy adjusts the distribution of subdomains with the same classes closer together in the feature space. \hyperref[subdomain]{Fig. \ref{subdomain}} depicts a comparison of the DA technique and the subdomain adaption method. The relationship between the data in these two domains must be utilized to split the source and target domains into multiple subdomains containing data of the same class. Nevertheless, since there are no labels in the target domain in unsupervised learning problems, the network output should be employed as the target domain data's pseudo-labels. As an outcome of the use of pseudo-labels, the source and target domains are divided into multiple subdomains $D_S^{(c)}$ and $D_T^{(c)}$ with probability distributions ${p^{(c)}}$ and ${q^{(c)}}$, respectively, where $c \in \left\{ {1,2,...,C} \right\}$ is the number of classes.
The suggested approach for subdomain adaptation is the LMMD technique \cite{zhu2020deep}, an extension of the MMD method that determines the distribution discrepancy of each same subdomain in different domains. The most commonly exploited nonparametric distance metric for DA is MMD, which measures the difference between two domains distribution in reproducing kernel Hilbert space (RKHS). MMD between ${D_S}$ and ${D_T}$ is described by the following relationship:
\begin{equation} 
\label{eqmmd}
d_{\mathcal{H}}(D_s,D_t)= {\| E_p[ \Phi(X^s) ]- E_q[ \Phi(X^t) ] \|}^2_{\mathcal{H}_{k}}
\end{equation}
In this context, $H$ is an RKHS, and $\Phi $ is a nonlinear mapping that converts data from ${D_S}$ and ${D_T}$ to RKHS feature space. To make computations easier, utilize the kernel characteristic $k$, which is defined by the relation $k\left( {{x^s},{x^t}} \right) = \left\langle {\Phi ({x^s}),\Phi ({x^t})} \right\rangle $, where $\left\langle {.,.} \right\rangle $ is the inner product of the vectors. As a result, an unbiased estimate of \hyperref[eqmmd]{Eq. \ref{eqmmd}} equals:
\begin{equation}
 \resizebox{\columnwidth}{!}{\begin{math}\begin{aligned}
&\hat{d}_{\mathcal{H}}(D_s,D_t)= \|  \frac{1}{n_s^2}\sum_{i=1}^{n_s} \sum_{j=1}^{n_s}k(x_i^s,x_j^s)+\frac{1}{n_t^2}\sum_{i=1}^{n_t} \sum_{j=1}^{n_t}k(x_i^t,x_j^t)\\
& -\frac{1}{n_sn_t}\sum_{i=1}^{n_s}\sum_{j=1}^{nt}k(x_i^s,x_j^t)\|^2_{{\mathcal{H}}_{k}}
\end{aligned}
\end{math}}
\end{equation}
In this case, ${n_s}$ and ${n_t}$ represent the number of source and target samples, respectively.
As previously stated, despite the MMD technique's effectiveness in measuring the distribution difference between two domains and playing an essential role in the regularization term of the loss function in some applications, there remains a need for a method to compute the distribution difference between each subdomain. The LMMD approach is employed in this study as generalized criteria of MMD to determine this discrepancy, which is defined as follows:
\begin{equation}
d_{\mathcal{H}}(D_s,D_t)= {\left \| E_p (c)[ \Phi(X^s) ]- E_q(c)[ \Phi(X^t) ]\right \|}^2_{{\mathcal{H}}_{k}}
\end{equation}
Each sample in each class is supposed to be allocated a weight $\mathcal{W}$. The relationship mentioned above will be as follows in this case:
\begin{equation}
\label{lmmd}
\resizebox{\columnwidth}{!}{\begin{math}\begin{aligned}
       L_{LMMD}=\hat{d}_{\mathcal{H}}(D_s,D_t)=\frac{1}{C}\sum_{m=1}^{c}\left \| \sum_{i=1}^{n_s}\mathcal{W} _i^{sm}\Phi(X_i^s) -\sum_{j=1}^{n_t}\mathcal{W} _i^{tm}\Phi(X_j^t) \right \|^2_{{\mathcal{H}}_{k}}
       \end{aligned}\end{math}
       }
\end{equation}
Where $\mathcal{W}_i^{sm}$ and $\mathcal{W}_j^{tm}$ represent the weight of $x_i^s$ and $x_j^t$ pertaining to class m, respectively.It should be noted that $\sum_{m=1}^{n_s}\mathcal{W}_i^{sm}=\sum_{m=1}^{n_t}\mathcal{W}_j^{tm}=1$ and $\mathcal{W}_i^m$ for the sample $x_i$ is calculated as:
\begin{equation}
 \mathcal{W}_i^m=\frac{\mathcal{Y}_{im}}{\sum_{xi,yi}^{}\mathcal{Y}_{jm}}   
\end{equation}
$\mathcal{W}_i^m$ requires ${y_i^s}$ and $y_i^t$ values, and as previously stated, a ${D_S}$ data label is provided, but $y_i^t$ must be determined using a pseudo-label.

$\mathcal{W}_j^{tm}$ can be calculated by obtaining $\hat{\mathcal{Y}}_i^t$ for each sample. \hyperref[lmmd]{Eq. \ref{lmmd}} can be modified to adjust the source and target domain features in the $l-th$ layer where $z^l$ is the lth layer activation of L layer.:
\begin{equation}
\label{lmmdf}
 \resizebox{\columnwidth}{!}{\begin{math}\begin{aligned} &\hat{d}_H(D_s,D_t)=\frac{1}{c}\sum_{m=1}^{c}[\ \sum_{i=1}^{n_s} \sum_{i=1}^{n_s}\mathcal{W}_i^{sm}\mathcal{W}_j^{sm}k(z_i^{sl},z_j^{sl})\\+
 &\sum_{i=1}^{n_t} \sum_{i=1}^{n_t}\mathcal{W}_i^{tm}\mathcal{W}_j^{tm}k(z_i^{tl},z_j^{tl})-2\sum_{i=1}^{n_s}\sum_{j=1}^{n_t}\mathcal{W}_i^{sm}\mathcal{W}_j^{tm}k(z_i^{sl},z_j^{st})]\ \end{aligned}\end{math}}
\end{equation}
\subsection{Domain Adversarial Network}
The adversarial domain adaptation network is one of the most famous fault diagnosis architectures. Applying This technique in unsupervised fault diagnosis has increased due to its powerful ability to reduce the distribution discrepancy between data in different domains \cite{huang2020deep}. The architecture contains two incorporating networks; a feature extractor and a domain discriminator network. On the one hand, the Feature extractor network aims to extract knowledgeable and discriminative features from the source and target domain. On the other hand, A domain discriminator network distinguishes whether extracted features come from the target or source domain. It inherits a binary classification algorithm to learn a logistic regressor for mapping features between [0,1]. Robust knowledge of source features maps into approximate target maps by fooling the feature extractor with a reverse gradient layer. Consequently, a classifier can use the knowledge to define unlabeled fault types in the target domain \cite{yu2020conditional}. As a final point, The Optimization objective as a binary cross entropy loss of the adversarial domain network is shown as follows:

\begin{equation}
\label{advformula}
\begin{aligned}
\mathcal{L}_{d}\left(x^{s}, x^{t}\right)=&-\mathbb{E}_{x^{\prime} \in \mathcal{D}_{x}}\left[\log D\left(G\left(x^{s}\right)\right)\right] \\
&-\mathbb{E}_{x^{\prime} \in \mathcal{D}_{t}}\left[\log \left(1-D\left(G\left(x^{t}\right)\right)\right)\right]
\end{aligned}
\end{equation}
where G(.) is the feature extractor, and D(.) accepts 0 or 1 as an input.

\begin{figure*}[t]
    \centering
    \includegraphics[width=\textwidth]{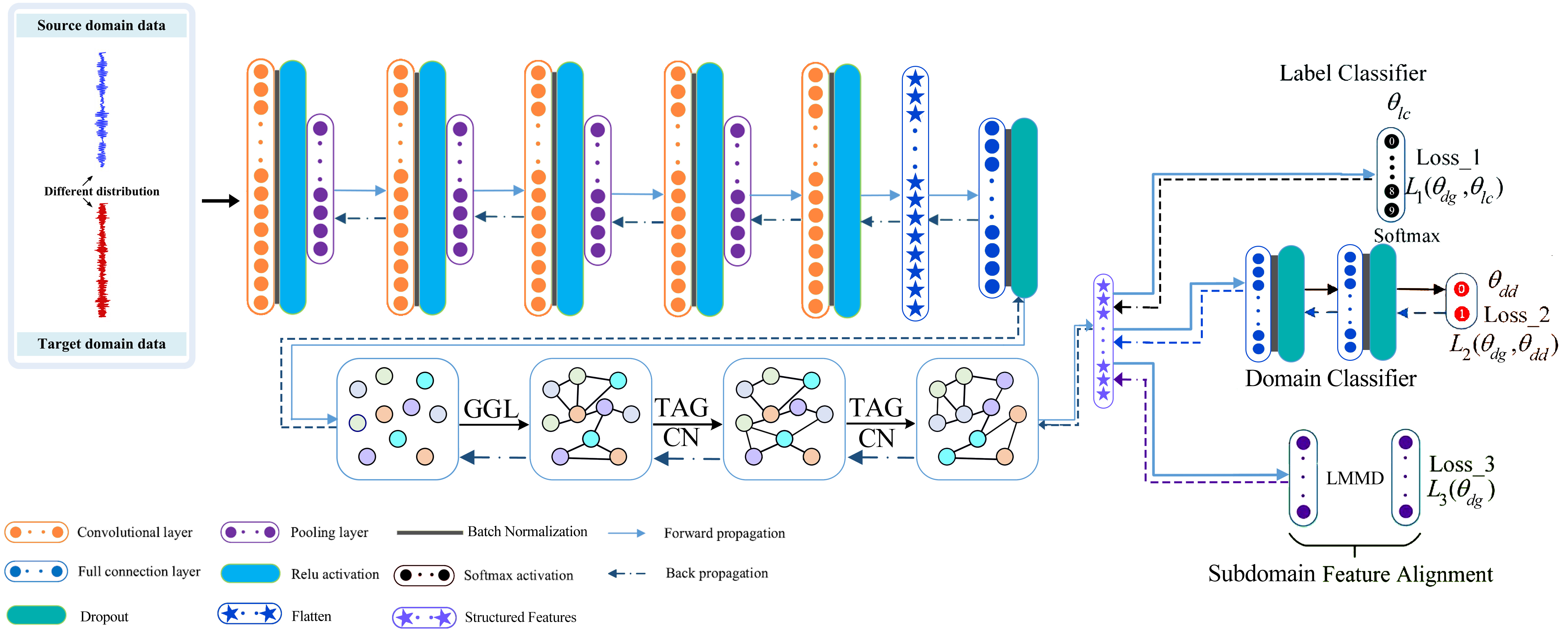}
    \caption{Proposed method}
\label{proposed}
    \end{figure*}
\section{PROPOSED METHOD} \label{sec3}
In this section, we discuss the detailed architecture of the proposed method in Fig. \ref{proposed}. As shown in Table \ref{structure}, We divide our model into graph feature extraction, domain adaptation, and classifier networks, which is detailed below:

\subsection{Graph Feature Extraction}
We employed a five-layer CNN with a wide kernel in the first part of the  feature extraction network for longer dependencies. As we proceed deeper into the layers, the kernel size decrease, improving local graph feature extraction and feature representation. Furthermore, using a wide kernel rather than the small one in the beginning layers causes high-frequency environmental noise to be repressed in input data to have a more robust network in the classification task \cite{zhang2018deep}. Directly after the convolution operation, the batch normalization technique is used to accelerate the network's training  and decrease the shift of internal covariance \cite{bt}. Rectified linear unit (ReLU) is utilized as an activation function to improve the representation ability and learn the complex pattern in the data \cite{zhang2018deep}. We used two kinds of pooling layers to reduce the network's parameters. Accordingly, the max-pooling layers directly after Relu activation layer and adaptive max-pooling are employed at the top of the CNN for a given reduced fixed-size output dimension. In the next stage, we reduced the dimensions of the feature vector by a dense layer $(FC1(.)$) with 256 neurons to have better robust feature representation. As the input of TAGCN is structured, it is necessary to transform unstructured features to structured ones, the output of the CNN network passes to a graph generation layer (GGL)\cite{li2021domain} for producing structured graph data, and then TAGCN is applied to structure  features. Each feature vector is specified as a node whose feature vectors come from the output of a dense layer, and an adjacency matrix is defined by the multiplication of the features vector and its transpose, as shown below:

\begin{table}[t]
\caption{Structures of Proposed Method}
\label{structure}
\Huge
\resizebox{\columnwidth}{!}{

\begin{tabular}{@{}cccccc@{}}
\toprule
\textbf{Network}                     & \textbf{Layer}                       & \textbf{Kernels Size /Stride/Filter Number} & \textbf{Output size} & \textbf{Pooling Size} & \textbf{Padding} \\ \midrule
\multirow{10}{*}{ Graph Feature Extractor}  & Conv1D, BN,  Relu, Max Pool          & 128/1*1/16                                  & N*1024*16            & 2*2                   & Yes              \\
                                     & Conv1D, BN,  Relu, Max Pool          & 64/1*1/32                                   & N*521*32             & 2*2                   & Yes              \\
                                     & Conv1D, BN,  Relu, Max Pool          & 32/1*1/64                                   & N*256*64             & 2*2                   & Yes              \\
                                     & Conv1D, BN,  Relu, Max Pool          & 16/1*1/128                                  & N*128*128            & 2*2                   & Yes              \\
                                     & Conv1D, BN,  Relu, Adaptive Max Pool & 3/1*1/128                                   & N*4*128              & 32*32                 & Yes              \\
                                     & FC1, Relu, Dropout 0.5               & 256 neurons                                 & N*256*1              &                       & NO               \\
                                     & GGL, Dropout                    & None                                 & N*N                  &                       & NO               \\
                                     & TGACN, BN                            & 128 neurons                                 & N*128                &                       & NO               \\
                                     & TGACN, BN                            & 256 neurons                                 & N*256                &                       & NO                                        \\ \midrule
\multirow{2}{*}{Domian Discrimnetor} & FC2, Relu, Dropout 0.5               & 128 neurons                                 & N*128                &                       & NO               \\
                                     & FC3,Relu, Dropout 0.5                & 128 neurons                                 & N*128                &                       & NO               \\
                                     & FC4, Sigmoid                         & 1 neurons                                   & N*1                  &                       & NO               \\ \midrule
Classifier                           & FC5, Softmax                         & Number of Fault types neurons                                  & N*10                 &                       & NO               \\ \bottomrule
\end{tabular}}

\end{table}
\begin{equation}
\label{CNNF}
X=CNN(Input \vspace{0.1cm} Data)
\end{equation}
\begin{equation}
\hat{X}=FC1(X)
\end{equation}
\begin{equation}
A=\mathcal{N}(\hat{X}\hat{X}^T)
\end{equation}
Where $\mathcal{N(.)}$ represents the normalization function. It is beneficial to make the adjacency matrix sparse to avoid computational costs. Therefore, as detailed below, K(.) returns the k-largest elements of the given adjacency matrix. Finally, Eq. \ref{TOPK} delivers a sparse adjacency matrix by indexing the top-k largest values of $A$ at row-wise. 
\begin{equation}
\label{TOPK}
\hat{A}=Top-K(A)    
\end{equation}
The generated graph will be fed through graph convolution layers in the next stage, extracting node features and aggregating neighbors' information. For better feature representation and robust aligned structured features, we used two TAGCN layers with the number of hops of two  ($K=2$). The output of each layer directly is connected to graph batch normalization layers as described in below:
\begin{equation}
\mathbf{x}= TAGCN(\hat{X},\hat{A},K)
\end{equation}
\begin{equation}
\label{BATCHN}
\mathbf{x}_{i}^{\prime}=\frac{\mathbf{x}-\mathrm{E}[\mathbf{x}]}{\sqrt{\operatorname{Var}[\mathbf{x}]+\epsilon}} \odot \gamma+\beta
\end{equation}
Where $\mathbf{x}_{i}^{\prime}$ shows normalized structured features; $E(.)$  and $Var(.)$ are expectation and variance functions, respectively; $\gamma$ and $\beta$ are trainable parameters; $\epsilon$ is added for numerical stability. To sum up, assuming Eq. \ref{CNNF} to Eq. \ref{BATCHN}, we considered CNN and GCNN network as a graph feature extractor network which is indicated by $G(.)$, where it takes input data and returns structured features.

\subsection{Domain Adaptation Networks}
As stated, we used two separate networks for aligning target and source features and reducing distribution between features in latent space, which are discussed in two following subsections:

\subsubsection{Domain Discriminator}
Adversarial domain learning tries to identify the label of extracted features supplied to the domain discriminator, as been described in Section \ref{sec2}. For producing invariant features in the target domain, reducing the loss computed inversion gradient by gradient reversal layer (GRL) in Eq. \ref{advformula}. We used FC2 and FC3 as a backbone of the domain discriminator network to produce more robust invariant latent space features.  As a result, the classifier will adopt either health states from the target domain or the source domain.

\subsubsection{LMMD Loss}
It is used at the top of the graph feature extractor part to reduce the distribution between extracted structured features. As a non-parametric technique, LMMD aims to match the distribution of source and target features and reduce the discrepancy distribution of relevant subdomains by integrating deep feature adaptation and feature learning. As been detailed in Eq. \ref{lmmdf}, the radial basis function (RBF) kernel is chosen as a kernel.
\subsection{classification layer}
We used a fully connected layer whose number of neurons equals the number of health conditions. The Softmax classifier is adopted as an activation function. Another usage is to produce pseudo labels in the target domain for calculating LMMD loss. By defining $H_{i}^{s},H_{j}^{s}=G(x_{i}^{s},x_{j}^{t})$, the objective function of classification $L_{C}$  is defined as:
\begin{equation}
\hat{y}_{j}^{\mathrm{t}}=\left\{\begin{array}{cc}
1 & \text { if } j=\underset{j}{\arg \max } H j^{\text {Softmax, } t} \\
0 & \text { otherwise }
\end{array}\right.
\end{equation}
\begin{equation}
\hat{y}_{i}^{\mathrm{s}}=\left\{\begin{array}{cc}
1 & \text { if } i=\underset{i}{\arg \max } H i^{\text {Softmax, } s} \\
0 & \text { otherwise }
\end{array}\right.
\end{equation}

\begin{equation}
\label{los3}
L_{C}\left(H_{i}^{s}, y_{i}^{s}\right)= E_{\left(x^{s}, s\right)-D} L\left(\hat{y}_{i}^{\mathrm{s}}), y_{i}^{s}\right)
\end{equation}
Where E(.) is considered mathematical expectation; $\hat{y}_{i}$ and $\hat{y}_{j}$ are generated logits by classification layer, respectively.
\begin{table}[t]
\caption{Algorithm of detailed DSAGCN Method}
\Huge{   
\resizebox{\columnwidth}{!}{
\begin{tabular}{@{}lllll@{}}
\toprule
\multicolumn{5}{l}{\textbf{Algorithm:  DSAGCN}}                                                                                                                                                                                                                                                                                                                                                                                                                                                                                                                                                                                                                                                 \\ \midrule
\multicolumn{5}{l}{\textbf{Require}: Raw Input Data, Learning Rate ($\eta$), Trade-off  Parameters($\mu$ , $\beta$)}                                                                                                                                                                                                                                                                                                                                                                                                                                                                                                                                                                 \\ \midrule
\multicolumn{1}{c}{1.}                  & \multicolumn{4}{l}{Define labeled source and unlabeled target dataset through pre-processing raw data}                                                                                                                                                                                                                                                                                                                                                                                                                                                                                                                                                                \\ \midrule
\multicolumn{1}{c}{2.}                  & \multicolumn{4}{l}{Initialize model's weights Using Xaviar initialization}                                                                                                                                                                                                                                                                                                                                                                                                                                                                                                                                                                     \\ \midrule
\multicolumn{1}{c}{3}                   & \multicolumn{4}{l}{Consider Input  $\{{x}{i}^{\mathrm{s}},{y}{i}^{\mathrm{s}}\} , \{{x}_{j}^{\mathrm{t}}\}$}                                                                                                                                                                                                                                                                                                                                                                                                                                                                                                                                                                                      \\ \midrule
\multicolumn{5}{l}{For $i,j = 1,......, n $ do :}                                                                                                                                                                                                                                                                                                                                                                                                                                                                                                                                                                                                                                          \\ \midrule
\multicolumn{1}{c}{4.}                  & \multicolumn{4}{l}{\begin{tabular}[c]{@{}l@{}}Graph Feature Extractor $\xleftarrow[]{}$ $G(.)$ \\ \\ $(H_{i}^{s}, H_{j}^{t}) \xleftarrow[]{} G({\{x_{i}^{s}, x_{j}^{t}\}_{i,j}^{n_s = n_t =n}})$\end{tabular}}                                                                                                                                                                                                                                                                                                                                                                                                    \\ \midrule
\multicolumn{1}{c}{5.}                                     & \multicolumn{4}{l}{\begin{tabular}[c]{@{}l@{}}Generate target pesudo labels :\\ \\ $\hat{{y}}_{j}^{\mathrm{t}}= \left\{\begin{array}{llll}1 & \text { if } & j=\underset{j}{\arg \max } {H}{j}^{\mathrm{Softmax}, \mathrm{t}} \\ 0 & & \text { otherwise }\end{array}\right.$\end{tabular}}                                                                    \\ \midrule
\multicolumn{1}{c}{6.}                                     & \multicolumn{4}{l}{\begin{tabular}[c]{@{}l@{}} Define total loss of  the DSAGCN method in forward pass:\\ \\

$L_{total}(H_{i}^{s}, H_{j}^{t}, {y}_{i}^{\mathrm{s}},\hat{{y}}_{j}^{\mathrm{t}})$ $\xleftarrow[]{}$ $L_{c}$($H_{i}^{s}, H_{j}^{t}$) + $\mu$ $L_{d}(H_{i}^{s}, H_{j}^{t})$ + $\beta$  $L_{Lmmd}(H_{s},H_{t},{y}_{i}^{\mathrm{s}},\hat{{y}}_{j}^{\mathrm{t}})$

\end{tabular}}                                                                                                                                                                                                                                                                                                                                                                                                                                                  \\ \midrule
\multicolumn{1}{c}{7.}                                      & \multicolumn{4}{l}{\begin{tabular}[c]{@{}l@{}} update weights of the DSAGCN model using Backpropagation algorithm:\\ $\theta_{g} \xleftarrow[]{} \theta_{g} - \eta \times \frac{\partial L_{Total}}{\partial \theta_{g}}$ 
\\ \\ $\theta_{d} \xleftarrow[]{} \theta_{d} - \eta \times \frac{\partial L_{d}}{\partial \theta_{d}}$ \\ \\ $\theta_{c} \xleftarrow[]{} \theta_{c} - \eta \times \frac{\partial L_{L_{c}}}{\partial \theta_{c}}$
\\  \\

\end{tabular}} \\ \midrule
\multicolumn{5}{l}{Until  $\theta_{c}$,  $\theta_{d}$, $\theta_{g}$  converge }                                                                                                                                                                                                                                                                                                                                                                                                                                                                                                                                                         \\ \bottomrule
\end{tabular}}}
\label{psoudo}
\end{table}

\subsection{Objective Function}
Considering classification loss (Eq. \ref{los3}), structured subdomain loss (Eq. \ref{lmmd}), and adversarial loss (Eq. \ref{advformula} )total loss of DSAGCN method with combining the defined three-loss described as follows:
\begin{equation} \resizebox{\columnwidth}{!}{\begin{math}\begin{aligned}
L_{\text {total }}\left(H_{i}^{s}, H_{i}^{t}, y_{i}^{s}, \hat{y}_{j}^{t}\right) = L_{c}\left(H_{i}^{s}, H_{i}^{t}\right)+\mu L_{d}\left(H_{i}^{s}, H_{j}^{t}\right)\\
+\beta L_{L m m d}\left(H_{s}, H_{t}, y_{i}^{s}, \hat{y}_{j}^{t}\right) \end{aligned}\end{math}}
\end{equation}
Where $\mu$ and $\beta$ are trade-off hyper-parameters. The algorithm of the proposed DSAGCN is summarised in the table \ref{psoudo}. It is shown that after producing total loss by network parameters of each network is updated with back-propagation operation and will be converged, which are detailed as below:
\begin{equation}
\begin{aligned}
&\theta_{g}=\theta_{g}-\eta \times \frac{\partial L_{\text {Total }}}{\partial \theta_{g}}\\
&\theta_{d}=\theta_{d}-\eta \times \frac{\partial L_{d}}{\partial \theta_{d}}\\
&\theta_{c}=\theta_{c}-\eta \times \frac{\partial L_{c}}{\partial \theta_{c}}
\end{aligned}
\end{equation}
Where $\theta_{g}$, $\theta_{d}$, and $\theta_{c}$ are parameters of graph feature extractor, domain discriminator, and classification layer, respectively; $\partial$ denotes partial derivative function; $\eta$ shows learning rate. 
\section{Experiments} \label{sec4}
The effectiveness of the proposed model is investigated in a variety of ways for two well-known datasets, including the CWRU \cite{loparo2003bearing} and Paderborn \cite{lessmeier2016condition} bearing datasets, in order to assess the validity of the proposed DSAGCN method in diagnosing bearing faults under various operating conditions.
\subsection{implementation details}
The Xavier initializer \cite{datta2020survey} is used during the training phase to set the DSAGCN method parameters
The initial learning rate of DSAGCN is set at 0.001. The Adam optimization technique \cite{kingma2014adam} is employed to optimize the parameters, and each batch has a length of 128. The optimal values of the trade-off parameters $\beta$ and $\mu$ are selected as 0.5 and 1, respectively. In addition, The degree of polynomial filter in the graph used in the DSAGCN method is considered equal to 2. Each experiment is repeated ten times to decrease the results' randomness. Fault diagnosis's average accuracy is used as assessment criteria.

\subsection{compared approaches}
To demonstrate the superiority of the proposed DSAGCN method over existing techniques, ten comparative methods including SVM, CNN, JDA \cite{long2013transfer}, CORAL \cite{sun2016return}, DANN \cite{ajakan2014domain}, unsupervised domain adaptation convolution neural network (UDACNN), Baseline, graph convolution maximum mean discrepancy (GC-MMD), graph convolution multi-kernel maximum mean discrepancy (GC-MKMMD), and graph convolution CORAL (GC-CORAL) have been implemented. All of these techniques use the same hyperparameters and settings as the DSAGCN method. These methods can be divided into the following categories:
\begin{enumerate}
\item \textbf{Traditional approaches}: SVM technique with radial basis function (RBF) kernel is used to evaluate the effectiveness of the proposed method compared to the traditional supervised method. Six features are chosen as the SVM model’s input, similar to \cite{xu2020transfer}. The SVM model is trained using extracted features from the labeled source domain, and the model is tested with unlabeled data in the target domain.
\item \textbf{DL-based approaches}:  The CNN network of the proposed graph feature extraction plus the classification layer is used as a CNN model compared with the proposed method in this study. Labeled source domain is utilized for CNN model training, whereas unlabeled target domain data is used exclusively for model testing. This model does not involve domain adaptation, and the only loss function is cross-entropy.
\item \textbf{DA-based techniques}:
Four domain adaptation approaches, including JDA, CORAL, DANN, and UDACNN, were utilized to demonstrate the superiority of the proposed method. 
The structure of the UDACNN approach is similar to the CNN model. Also, it incorporates two domain adaptation modules, including the domain discriminator and LMMD, similar to the DSAGCN method. The only difference between the UDACNN and DSAGNN models is the lack of a graph; all other variables and hyperparameters are the same.

\item \textbf{Graph-based techniques}: We considered either the backbone of DSAGCN  method domain with discriminator network in  GC-MKMMD, GC-MMD and GC-CORAL 
or the Baseline method without them to compare the role of structure subdomain adaptation with other domain adaptation with non-parametric loss in our proposed method.
As previously stated, the baseline model includes the feature extraction and classification layer parts, but unlike the CNN model, it also includes a graph. In addition, unlike DSAGCN, it does not employ any domain adaptation modules. In  GC-MKMMD, GC-MMD, and GC-CORAL, the adversarial domain discriminator loss function and the cross-entropy loss function are employed, and Their difference is in the third loss function. In contrast to DSAGCN, the MMD loss method is applied in GC-MMD instead of LMMD. Also, for GC-MKMMD and GC-CORAL methods, the multi-Gaussian kernels with a mixture of five different bandwidths $\{0.001,0.01,1,10,100\}$ and the CORAL loss function are utilized, respectively. These three techniques aim to decrease the distribution difference between domains by applying three loss functions simultaneously and improving classification accuracy.

\end{enumerate}
 \begin{figure}[t]
\label{cwru}
    \centering
    \includegraphics[width=\columnwidth,height=3.5cm]{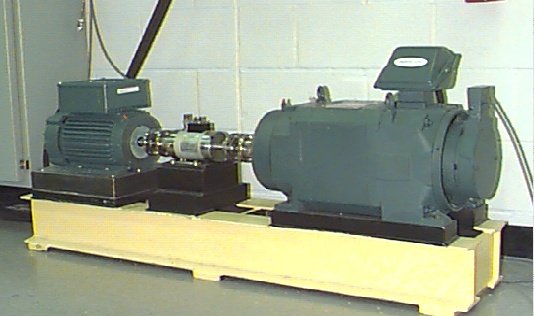}
    \caption{The experimental setup of CWRU bearing vibration dataset}
    \label{cwru}
    \end{figure}

\begin{table}[]
\caption{description of the CWRU bearing dataset}
\label{dataset_cwru}
\begin{tabular}{cccc}
\hline
Working Condition & Load & Types of Fault & Fault Diameter (mils) \\ \hline
A                 & 0 hp & N,IRF,ORF,RF   & 7,14,21                   \\ \hline
B                 & 1 hp & N,IRF,ORF,RF   & 7,14,21                   \\ \hline
C                 & 2 hp & N,IRF,ORF,RF   & 7,14,21                   \\ \hline
D                 & 3 hp & N,IRF,ORF,RF   & 7,14,21                   \\ \hline
\end{tabular}
\label{dataset_cwru}
\end{table}
\subsection{Case I: CWRU Expriment}
\subsubsection{data description}
The CWRU bearing vibration dataset is utilized to assess the efficiency of the suggested DSAGCN method, which the test platform depicted in Figure 4. In this work, the CWRU bearing vibration dataset is utilized to assess the efficiency of the suggested DSAGCN method, which the test platform depicted in Fig. \ref{cwru}. Vibration data are obtained using an accelerometer with a sampling rate of 12kHz located at the motor’s drive end. Data are gathered under four different operating conditions caused by load changes ranging from 0 to 3 hp. This dataset considers four different bearing health modes:  Normal, Inner Race fault (IRF), Outer Race Fault (ORF), and Roller Fault (RF). Each of them has three different severity faults with diameters 7, 14, and 21 mils. Therefore, this dataset has ten different states of bearing health status, which is given in Table \ref{dataset_cwru}. Data augmentation is used to enhance the number of samples, which improves the performance of the training process. As a result, a sliding window with a length of 1024 data points and a time step of 475 is employed. Finally, there are 200 training and 50 test samples for each motor load after shuffling samples. There are 12 transfer learning tasks based on four different loads. Each task is divided into two parts: the source and the target. For instance, in task B→D, the labeled data from dataset D is used as the source domain, whereas unlabeled data from dataset B is used as the target domain.
The DSAGCN method is trained on the CWRU dataset across 100 epochs.
\begin{figure*}[ht]
    \centering
    \includegraphics[width=\textwidth]{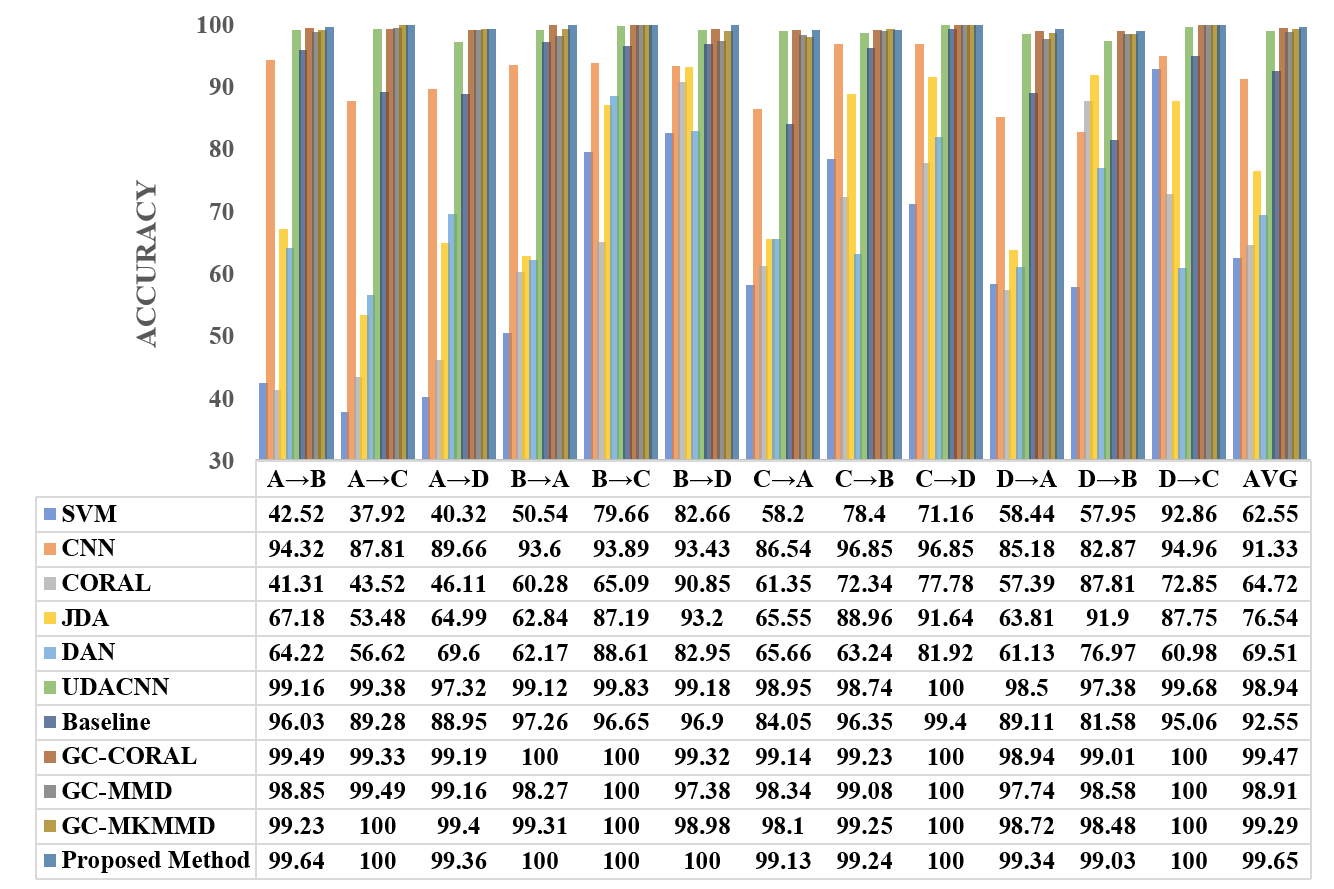}
    \caption{Diagnosis accuracy of different methods on CWRU bearing dataset under changing working conditions}
\label{cwru_result}
    \end{figure*}
\subsubsection{ CWRU bearing fault diagnosis Result and Discussion} 
\hyperref[cwru_result]{Fig. \ref{cwru_result}} depicts the simulation results for various tasks using the DSAGCN method and comparison approaches. The collected results can be categorized as follows:
\begin{figure}[t]
    \centering
    \includegraphics[width=\columnwidth]{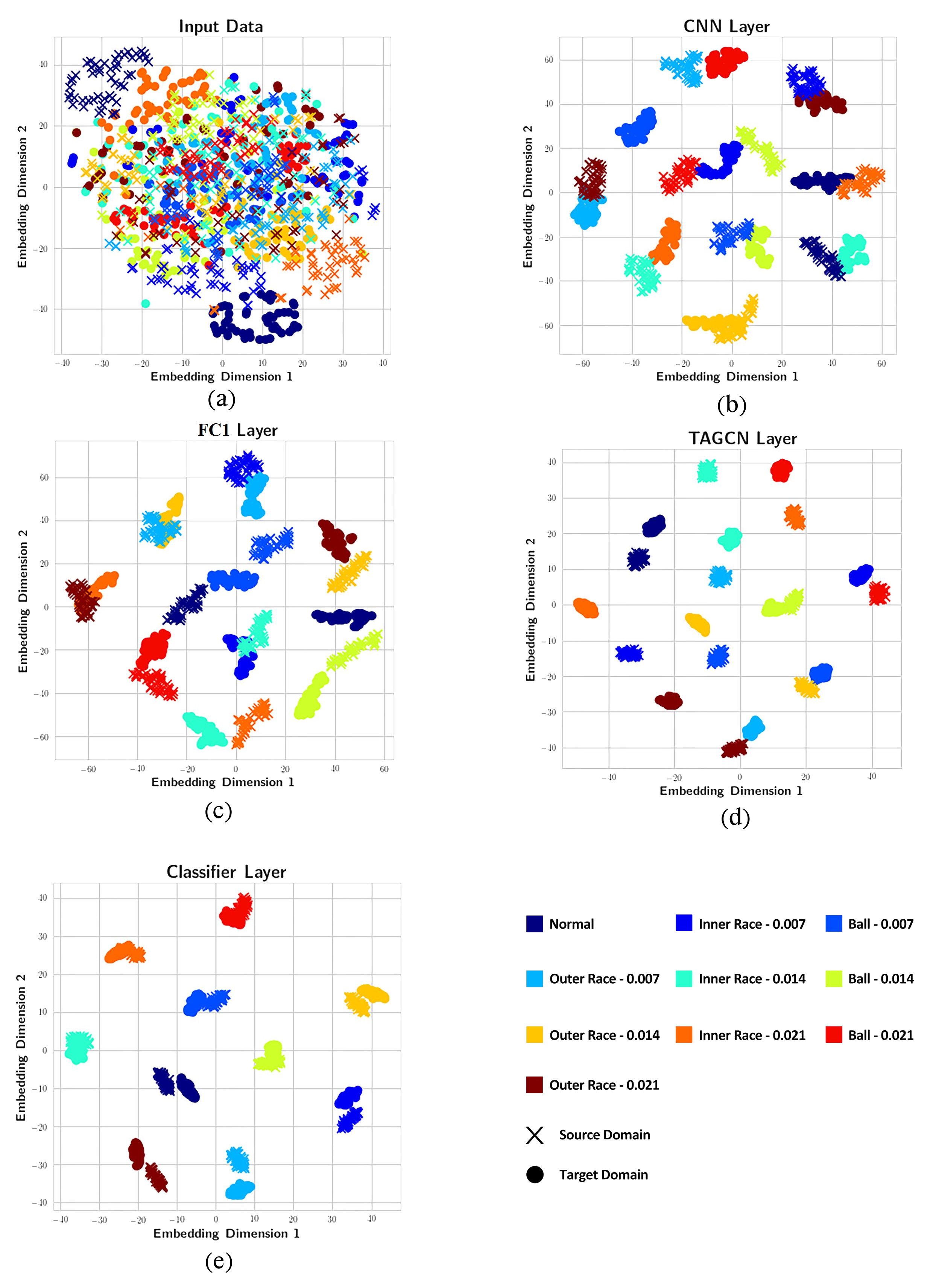}
    \caption{t-SNE feature visualization of proposed DSAGCN method for task $B\to D$  of CWRU dataset in different layers: (a) input layer, (b) CNN layer, (c) FC1 Layer, (d) TAGCN layer, (e) classifier layer(\%)}
\label{tsne_cwru1}
    \end{figure}
\begin{enumerate}
\item It is evident from the results that the DSAGCN method outperforms the other comparison methods in unlabelled fault diagnosis. The DSAGCN method achieves higher accuracy by utilizing the geometric structure of data in graphs, matching each structured subdomain using the LMMD algorithm, and minimizing the distribution discrepancy between domains using the adversarial loss function. As a result, because it covers all three types of essential information for the UDA technique, the DSAGCN method surpasses the three traditional, DL-based, and DA-based procedures. Furthermore, the DSAGCN method outperforms graph-based techniques in terms of accuracy, emphasizing the significance of deep structured subdomain adaptation methods.
\item In one of the most challenging tasks,B→D, the worst fault diagnosis accuracy is 57.95\%, while the best accuracy is related to the proposed DSAGCN method with 99.03\%. The DSAGCN method's fault diagnosis accuracy is 0.02\%, 1.65\%, and 16.16\%, greater than the best accuracy of the three categories, graph-based, DA-based, and DL-based. This high accuracy demonstrates the DSAGCN method's efficiency when there is a substantial distribution difference between domains.
\item In graph-based techniques, GC-CORAL, GC-MMD, and GC-MKMMD have higher average accuracy than DA-based methods, demonstrating the necessity of using hybrid DA tools and the importance of data geometry in approaching the distance between two domains. As can be observed, the geometric structure of the data is an essential aspect in improving the UDA's performance. Also, this group of approaches has been offered as an ablation study. In contrast to the DSAGCN approach, GC-MMD, GC-MKMMD, and GC-CORAL employ MMD, MKMMD, and CORAL instead of LMMD. The findings demonstrate the superiority of the suggested strategy and the effectiveness of utilizing LMMD.
\item Due to a lack of domain matching ability to decrease the distribution difference, the baseline technique has the lowest average accuracy compared to other graph-based methods and the suggested DSAGCN method. The baseline technique outperforms CNN in terms of average accuracy, indicating the efficacy of the graph and the geometric structure of the data in lowering the distribution disparity. The baseline approach is less accurate than the UDACNN, meaning that the geometric data structure is not powerful enough to alleviate the distribution difference without other DA techniques. Performance is when improved LMMD-based and adversarial domain adaptation combination techniques are used.
\item The performance of the DSAGCN method, graph-based, and domain-based techniques outperform the performance of the other two groups. The SVM approach does not perform as well as other methods since it does not use all of the key information of vibration data owing to reasons like the selection of hand-made features extracted as input, traditional machine learning networks' low performance in identifying complicated nonlinear relationships between data, and training the model only using Source data. When compared to other methods, it has the worst average accuracy. Although the CNN model outperforms the SVM model due to deep learning networks' capacity to analyze nonlinear connections and automated feature learning, it has lower classification accuracy than the graph-based and UDACNN methods, due to reasons such as discrepancies in distribution between training and test data, a lack of adequate tools to decrease distribution disparities, and a failure to pay attention to the varied geometric structure of training and test data. The accuracy of the CNN model demonstrates that the trained model in a given load does not have excellent diagnosis accuracy in classifying bearing faults in other loads. The features acquired in the primary layers of the CNN model are more generic and can be utilized for data with various distributions. However, as they progress to the end layers, the features become more particular, necessitating the employment of a tool such as DA to transmit information from the source domain to the target domain, decrease distribution discrepancies, and consequently enhance fault detection performance.
\end{enumerate}
\begin{figure}[t]
    \centering
    \includegraphics[width=\columnwidth]{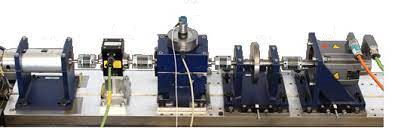}
    \caption{Experimental platform of Paderborn bearing dataset}
\label{pad}
    \end{figure}

\begin{table}[t]
\caption{description of the Paderborn bearing dataset}
\label{PADERBORN}
\resizebox{\columnwidth}{!}{%
\begin{tabular}{ccccc}
\hline
\multirow{2}{*}{Dataset} & \multirow{2}{*}{Faulty condition} & \multicolumn{3}{c}{Working Condition}                        \\ \cline{3-5} 
                         &                                   & Rotational Speed (rpm) & Load Torque (Nm) & Radial Force (N) \\ \hline
E                        & N,IRF,ORF                         & 900                    & 0.7              & 1000             \\ \hline
F                        & N,IRF,ORF                         & 1500                   & 0.1              & 1000             \\ \hline
G                        & N,IRF,ORF                         & 1500                   & 0.7              & 400              \\ \hline
H                        & N,IRF,ORF                         & 1500                   & 0.7              & 1000             \\ \hline
\end{tabular}%
}
\end{table}
The t-distributed stochastic neighbor embedding (t-SNE) approach \cite{van2008visualizing} is applied  to intuitively understand the suggested DSAGCN method's efficiency in reducing the difference in the distribution of learned features across two domains and aligning relevant subdomains with the same class in two domains. This method converts learned features with high dimensions to 2-D feature space.
Fig. \ref{tsne_cwru1} depicts the t-SNE representation of the $B\to D$ task as the most challenging diagnostic task in five-step in the DSAGCN method.
Fig. \ref{tsne_cwru1}-a exhibits the model's input data, which reveals that the data from various classes are mixed and not proper for classification. The output of the adaptive max-pooling layer, which represents the CNN part of the model, is shown in Fig. \ref{tsne_cwru1}-b. The distribution disparity between the source and target domains is visible in this figure. In fig. \ref{tsne_cwru1}-c with the inclusion of the FC1 layer in the continuation of the CNN section, the segregation performance is better than that of CNN, but the data of the distinct categories is still incorrectly classified. For example, IRF class data with a severity of 0.021 mils are not adequately distinguished from ORF class data with severity of 0.021 mils, which has a detrimental impact on fault diagnosis performance. The data of various classes are well segregated from each other in Fig. \ref{tsne_cwru1}-d, which visualizes the output of the TAGCN layer, although the disparity between the distributions between domains is substantial. Domain adaptation is employed to tackle this issue, as seen in Fig. \ref{tsne_cwru1}-e. This figure indicates the beneficial effect of subdomain adaptation on reducing distribution disparities and improving class separation.
\begin{figure*}[t]
    \centering
    \includegraphics[width=\textwidth]{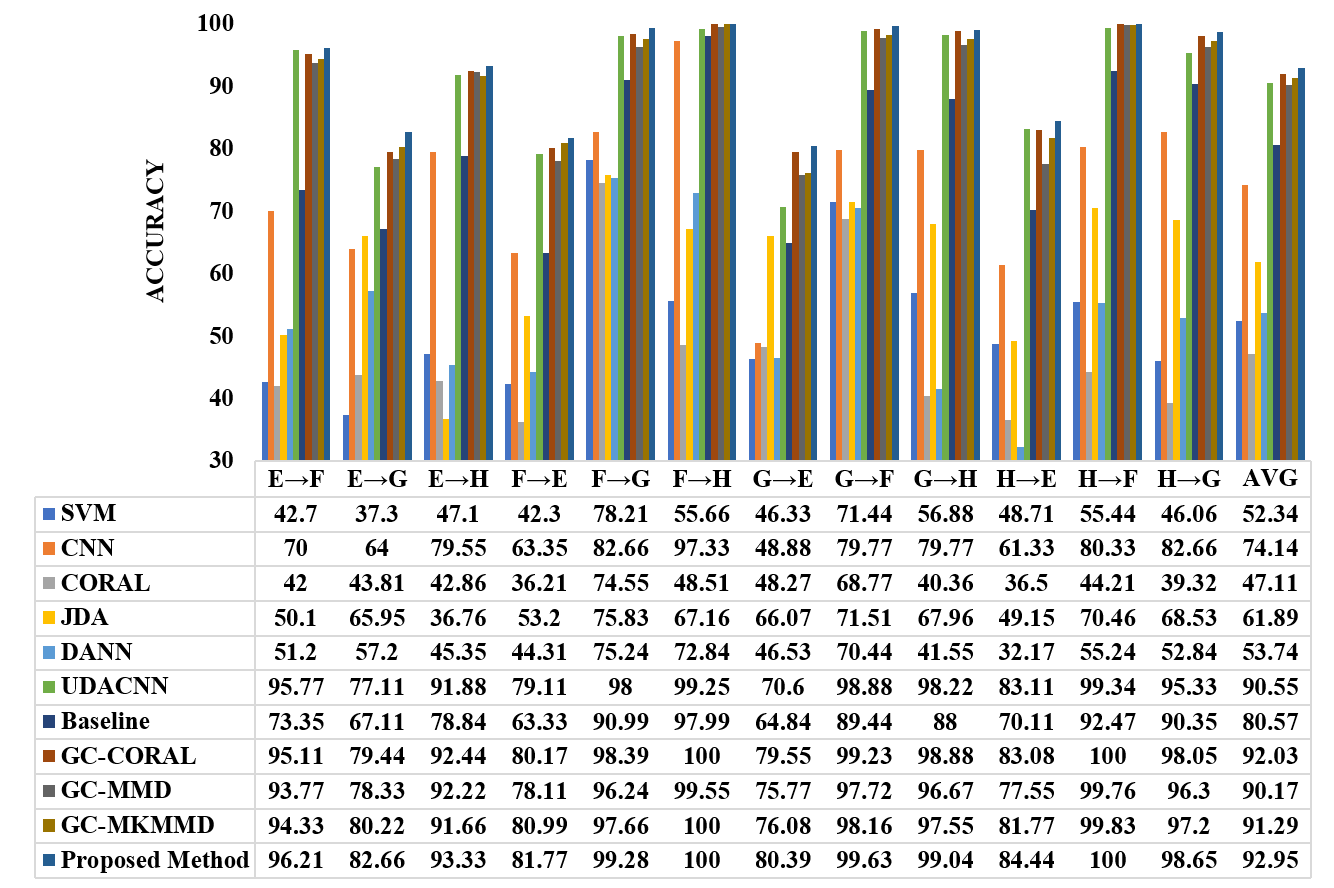}
    \caption{Diagnosis accuracy of different methods on Paderborn bearing dataset under changing working conditions}
\label{paderborn_result}
    \end{figure*}
\subsection{PU Expriment}
\subsubsection{data description}

The experimental data are gathered from the Paderborn university rig test\cite{lessmeier2016condition}, which included five electric motor components: a measuring shaft, a rolling bearing module, a flywheel, and a load motor. Fig. \ref{pad} depicts a modular test rig for this dataset. It considers three different bearing health conditions: Normal, IRF, and ORF. Two artificial and real damage are supposed to generate the faults types. Given that two distinct values for rotational speed, load torque and radial force, were examined in this study. Table \ref{PADERBORN} investigated four different working conditions to assess fault diagnosis performance under various operating conditions. Nine distinct classes K001, K003, K005, KA04, KA16, KA22, KI04, KI14, and KI16, are conducted according to\cite{lessmeier2016condition}, to assess the proposed method. All selected ORF and IRF faults are chosen from the types of real bearing damage generated by accelerated lifetime testing. An accelerometer sensor with a sampling frequency of 64 kHz is used to collect vibration data. A 1024 datapoint sliding window with no overlap is employed to divide the data. Finally, for each motor operating condition, 200 samples for the train and 50 samples for the test are provided. On the Paderborn dataset, the DSAGCN method is trained for 400 epochs.
\subsubsection{Paderborn diagnosis Result and Discussion}
Fig. \ref{paderborn_result} depicts the simulation results of the proposed DSAGCN method and comparison methods for the unsupervised fault diagnosis issue for the Paderborn dataset under different working conditions. The gathered information can be divided into the following categories:
\begin{enumerate} 
\item When evaluating the average accuracy of fault diagnosis, the suggested DSAGCN approach has the highest accuracy compared to other methods. The average accuracy of the DSAGCN method is 0.92\%, 24\%, 18.81\%, and 40.61\%, more accurate than the highest.
\item In one of the most challenging transfer tasks, G→E transfer, the DSAGCN method has 0.84\% and 34.06\% better accuracy than the maximum and minimum diagnosis accuracy in comparative methods, respectively. When there is a significant distribution gap among domains, this high accuracy indicates the effectiveness of the DSAGCN approach.
\item The explanation for reduced fault diagnosis accuracy in some tasks, such as F→E, compared to other tasks is a change in working conditions and more severe differences in distribution between domains. In four tasks, the average accuracy of the DSAGCN approach was less than 90\%.
\item
The mean accuracy of the DSAGCN method and other comparison methods for the Paderborn dataset is much lower than the results obtained for the CWRU dataset in Fig. \ref{cwru_result}. The suggested DSAGCN method’s average accuracy for the Paderborn dataset is 6.7\% lower than the CWRU dataset due to the Paderborn dataset’s greater complexity than the CWRU and the significant distribution differences between domains in the Paderborn tasks. In the CWRU dataset, only the load changes, and the rotational speed is relatively constant. In contrast, in the Paderborn dataset, there is the possibility of changing the rotational speed and radial force, which causes more differences between distributions and reduces the accuracy of fault diagnosis under different operating conditions. 
\end{enumerate}
\begin{figure}[t]
    \centering
    \includegraphics[width=\columnwidth]{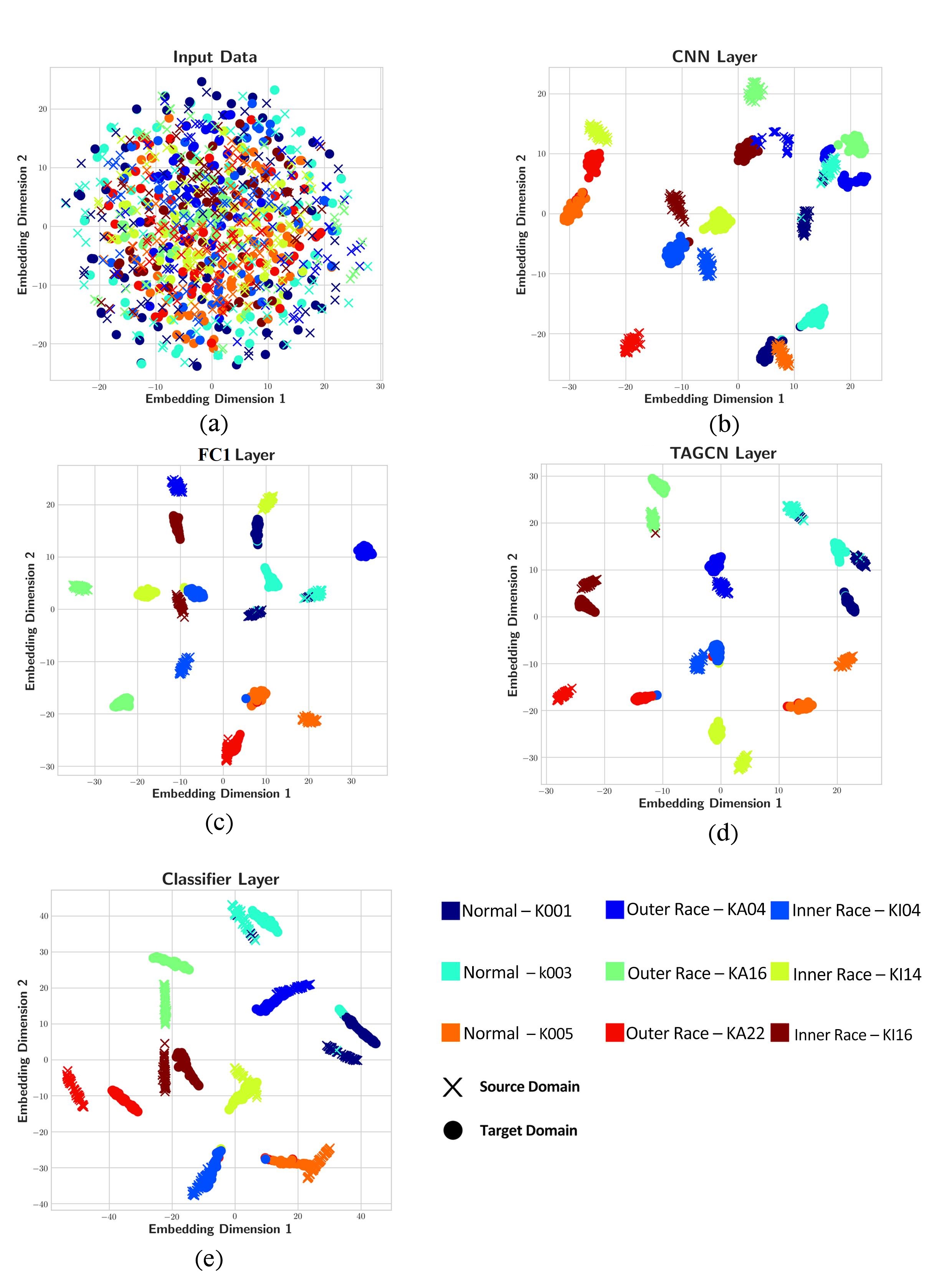}
    \caption{t-SNE feature visualization of proposed DSAGCN method for task H→G  of Paderborn dataset in different layers: (a) input layer, (b) CNN layer, (c) FC1 Layer, (d) TAGCN layer, (e) classifier layer.}
\label{tsne_pad}
    \end{figure}
In addition to analyzing numerical results to examine the effectiveness of the proposed method, this part provides a visual representation of t-SNE in five different layers, including input data, CNN layer, FC1 layer, TAGCN layer, and classifier layer. The H→G diagnostic task is randomly chosen, and the t-SNE is displayed in Fig. \ref{tsne_pad}. As can be seen, the gap between the source and target domains gets less as we get closer to the end layers, and data of the same class in the two domains get closer to each other in latent space. Due to domain adaptation modules in the end layers, the difference between the domains is reduced, and the domains are better aligned. However, a small amount of data are wrongly placed near another class. For example, some data of the k001 class are mixed up with the k003 class, lowering the accuracy of this task’s fault diagnosis. 
\subsection{Model Discussion} \label{discuss}
\subsubsection{degree of graph's polynomial filter }
One of the crucial parameters of the suggested approach is the value of $k$, which is the polynomial degree of the graph filter in the proposed DSAGCN model. In this subsection, $k$ values of \{1,2,4,5,10,25,100\} were considered, and the criterion of greatest accuracy and lowest training time was studied to achieve the ideal value of $k$. Table \ref{k_sen} illustrates the accuracy and training time for CWRU and PU datsets for various values of $k$ in a single epoch.  the B→D and H→G tasks are shown as selected task in Table\ref{k_sen}, respectively. The model was trained ten times for each value of $k$ to decrease randomness, and the average accuracy and training time were considered. According to Table \ref{k_sen}, 100\% accuracy is achieved for $k = 2,4,5,10$ to diagnose the fault in the CWRU dataset; however, as the value of $k$ increases, the training time increases owing to the increasing complexity and computational burden. Because of the rising complexity, diagnostic accuracy decreases dramatically for $k=50,25$. Table \ref{k_sen} demonstrates that the highest accuracy of fault diagnosis is obtained for $k=2$. As the value of $k$ increases, the accuracy declines substantially, reaching an accuracy of 16.25\% for $k=100$, which is less than the accuracy of all comparison approaches.  As a result, the optimal selection of the $k$ value in the suggested model is crucial. According to the results obtained on both datasets, $k=2$ is the best choice for the suggested model, with the maximum accuracy and a short training time.
\begin{table}[t]
\label{k_sen}
\caption{Investigation of polynomial degree selection of graph filter in diagnosis accuracy and training time of DSAGCN method for B→D and H→G tasks in two datasets}
\Huge
\resizebox{\columnwidth}{!}{%
\begin{tabular}{@{}cccccccccc@{}}
\toprule
\multirow{3}{*}{CWRU Dataset} & K             & 1     & 2     & 4     & 5     & 10    & 25   & 50    & 100   \\ \cmidrule(l){2-10} 
                              & Accuracy (\%) & 70.44 & \textbf{100}   & 100   & 100   & 100   & 99.8 & 83.4  & 66    \\ \cmidrule(l){2-10} 
                              & Time (S)      & 391   & \textbf{396}   & 427   & 444   & 536   & 815  & 1273  & 2176  \\ \midrule
\multirow{2}{*}{PU Dataset}   & Accuracy (\%) & 84.44 & \textbf{98.63} & 87.42 & 85.55 & 68.34 & 50.6 & 25.33 & 16.25 \\ \cmidrule(l){2-10} 
                              & Time (S)      & 1414  & \textbf{1458}  & 1523  & 1603  & 2018  & 3071 & 4772  & 6023  \\ \bottomrule
\end{tabular}}
\label{k_sen}
\end{table}
 \begin{figure*}[t]
    \centering
    \includegraphics[width=\textwidth]{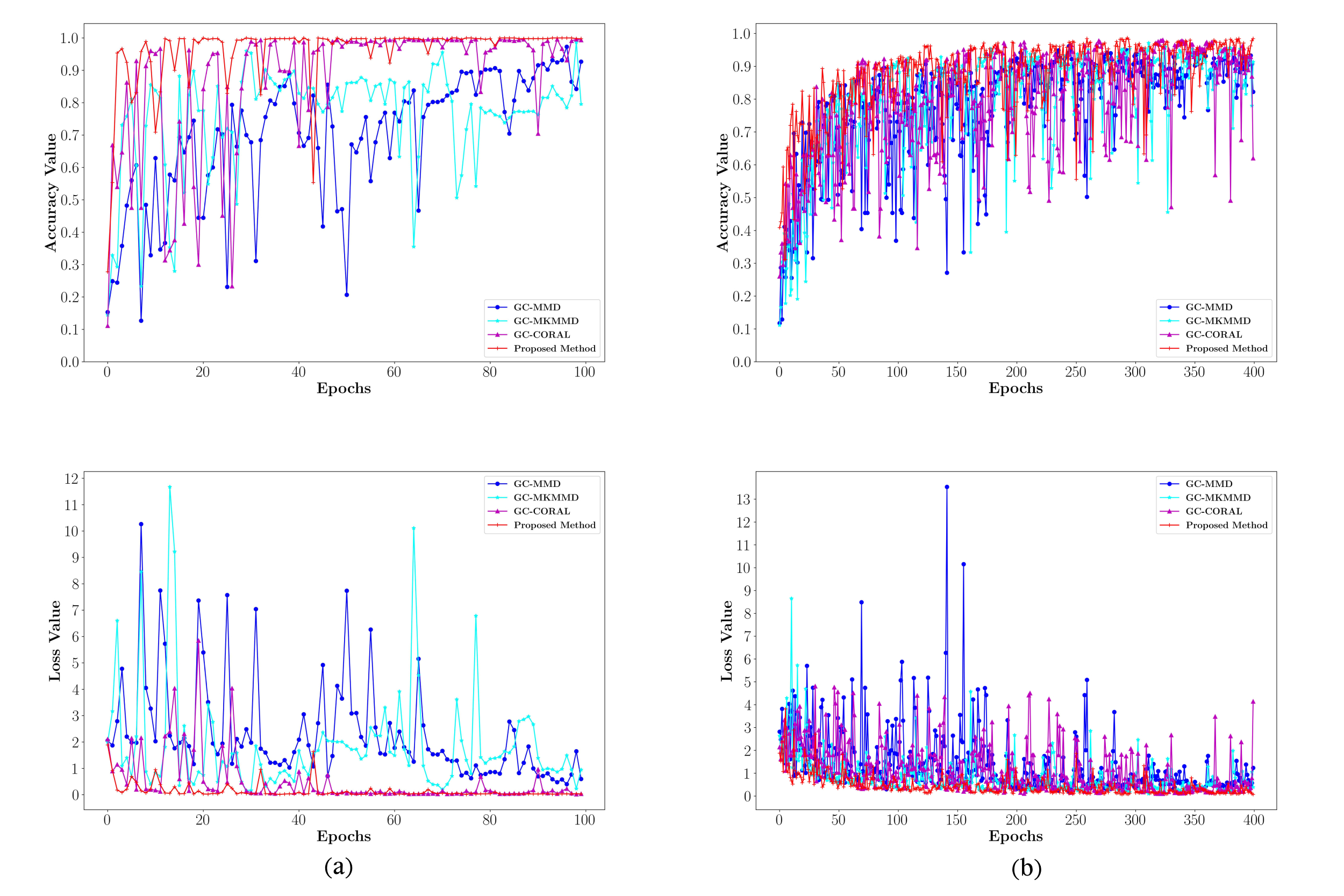}
    \caption{Convergence performance of the proposed model in terms of the total loss function and diagnosis accuracy; (a) task B→D of CWRU dataset (b) task H→G of PU dataset}
\label{converge}
    \end{figure*}
\begin{figure}[t]
    \centering
    \includegraphics[width=\columnwidth]{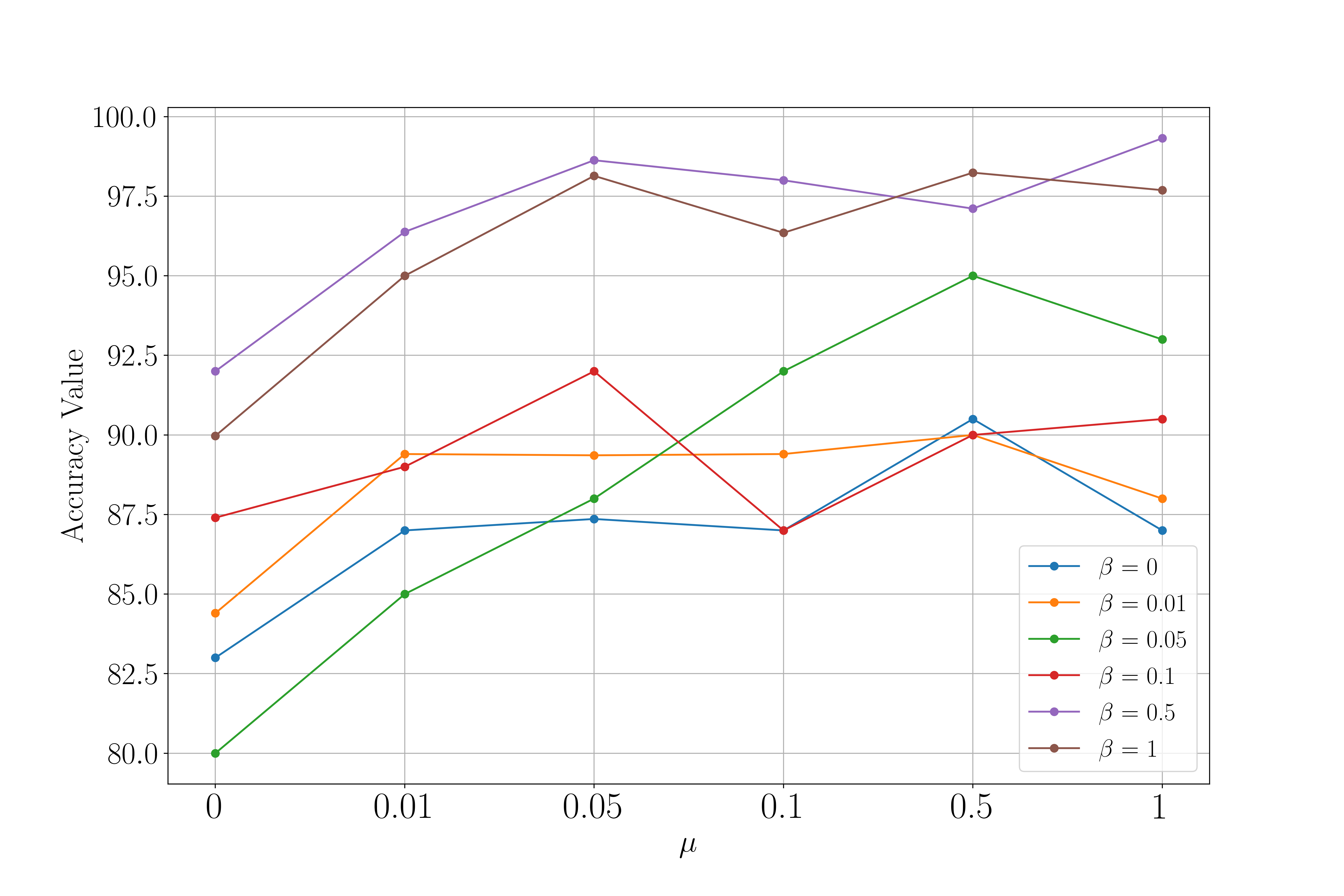}
    \caption{Comparison of fault diagnosis results in the B→D transfer task for different values of the loss function's penalty coefficients}
\label{hyperparameter}
    \end{figure}
\subsubsection{Convergence performance}
Fig. \ref{converge} depicts a representation of the total loss function and diagnostic accuracy per epoch in the B→D  and H→G tasks on test data for the CWRU dataset  and PU dataset, respectively. In this figure, the suggested DSAGCN method is compared to the three techniques GC-MMD, GC-MKMMD, and GC-CORAL to assess the effectiveness of the LMMD method. As can be observed, as the number of epochs rises, the total loss rate drops while the accuracy rate improves. Furthermore, the DSAGCN method converges to its lowest loss function faster than other approaches, indicating that this model is better and more efficient than other methods in terms of convergence. In terms of fluctuations, the DSAGCN method has the lowest fluctuations compared to other methods, and it has a smoother curve, demonstrating the approach's high stability. In conclusion, Fig. \ref{converge} demonstrates the superiority of the suggested approach and the perfect performance of the LMMD technique when contrasted to CORAL, MMD, and MKMMD.
\subsubsection{coefficient of the loss function:}
Six distinct values \{0,0.01,0.05,0.1,0.5,1\} were used as $\beta$ and $\mu$ coefficients to investigate the influence of the coefficients of two adversarial and LMMD Loss functions on the fault diagnosis outcomes. Fig. \ref{hyperparameter} shows an example of the B→D transfer task from the CWRU dataset. As can be observed, the model has high accuracy for different values of $\mu$ at $\beta$ = 0.5, $\beta$ = 1, indicating the strong inﬂuence of the LMMD approach. $\beta$= 0.5 and $\mu$ = 1 are the most optimal values of $\beta$ and $\mu$. For values of $\beta$= $\mu$ = 0, the domain adaptation modules have no impact, and the mode is equivalent to the baseline method.

\section{ conclusion} \label{sec6}
This paper presents a novel end-to-end DSAGCN approach and a new framework for bearing fault diagnosis under diverse operating conditions based on crucial information such as data geometry. Adversarial domain adaptation and structured subdomain adaptation are employed to decrease the distribution discrepancy. Also, the geometric structure of data is obtained using graph theory in this study.
The suggested model's graph, which has a lower computing complexity than the spectrum technique and no linear approximation in its calculation, considers the geometric structure of the data under different working conditions and aids in diagnosing bearing faults. On the other hand, the adversarial domain discriminator is used to align the domains, and the LMMD loss function is utilized to match subdomains with the same class.
Experiment results on the CWRU and Paderborn datasets demonstrate that the DSAGCN method outperforms other approaches in terms of fault diagnosis accuracy under various operating conditions.

\bibliographystyle{ieeetr}
\bibliography{main.bib}

\vfill

\end{document}